\documentclass[preprint2]{aastex62}
\pdfoutput=1 
\usepackage{amsmath,amstext}
\usepackage[T1]{fontenc}
\usepackage{apjfonts}
\usepackage[figure,figure*]{hypcap}
\usepackage{footmisc}
\newcommand{\angstrom}{\mbox{\normalfont\AA}}

\shorttitle{B-stars}
\shortauthors{Ramirez-Preciado V et al.}

\begin{document}

\title{SPECTRAL CLASSIFICATION OF B STARS: THE EMPIRICAL SEQUENCE USING SDSS-IV/APOGEE NEAR-IR DATA}

\author[0000-0002-4013-2716]{Valeria G. Ram\'irez-Preciado}
\affiliation{Instituto de Astronom\'ia, Universidad Nacional Aut\'onoma de M\'exico, Unidad Acad\'emica en Ensenada, Ensenada BC 22860 Mexico}
\author[0000-0002-1379-4204]{Alexandre Roman-Lopes}
\affiliation{Departament of Physics and Astronomy, Universidad de La Serena, Av. Juan Cisternas, 1200 North, La Serena, Chile}
\author[0000-0001-8600-4798]{Carlos G. Rom\'an-Z\'u\~niga}
\affiliation{Instituto de Astronom\'ia, Universidad Nacional Aut\'onoma de M\'exico, Unidad Acad\'emica en Ensenada, Ensenada BC 22860 Mexico}
\author[0000-0001-9797-5661]{Jes\'us Hern\'andez}
\affiliation{Instituto de Astronom\'ia, Universidad Nacional Aut\'onoma de M\'exico, Unidad Acad\'emica en Ensenada, Ensenada BC 22860 Mexico}
\author[0000-0002-1693-2721]{D. A. Garc\'ia-Hern\'andez}
\affiliation{Instituto de Astrof\'isica de Canarias (IAC), E-38205 La Laguna, Tenerife, Spain}
\affiliation{Universidad de La Laguna (ULL), Departamento de Astrof\'isica, E-38205 La Laguna, Tenerife, Spain.}
\author[0000-0002-3481-9052]{Keivan Stassun}
\affiliation{Vanderbilt University, Department of Physics \& Astronomy, 6301 Stevenson Center Lane, Nashville, TN 37235, USA}
\author[0000-0003-1479-3059]{Guy S. Stringfellow}
\affiliation{Department of Astrophysical and Planetary Sciences, Center for Astrophysics and Space Astronomy, University of Colorado, 389 UCB Boulder Colorado, USA}
\author[0000-0001-6072-9344]{Jinyoung Serena Kim}
\affiliation{Steward Observatory, University of Arizona 933 N. Cherry Avenue, Tucson, AZ 85721-0065, USA}

\begin{abstract}
We present a semi-empirical spectral classification scheme for normal B-type stars using near-infrared spectra (1.5-1.7 $\mu$m) from the SDSS APOGEE2-N DR14 database. The main motivation for working with B-type stars is their importance in the evolution of young stellar clusters, however we also take advantage of having a numerous sample (316 stars) of B-type star candidates in APOGEE2-N, for which we also have optical (3600-9100 \angstrom) counterparts from the LAMOST survey. By first obtaining an accurate spectral classification of the sources using the LAMOST DR3 spectra and the canonical spectral classification scheme \citep{Gray2009}, we found a linear relation between optical spectral types and the equivalent widths of the hydrogen lines of the Brackett series in the APOGEE2-N NIR spectra. This relation extends smoothly from a similar relation for O and early-B stars found by \citet{Roman2018}. This way, we obtain a catalog of B-type sources with features in both the optical and NIR, and a classification scheme refined down to one spectral sub-class. 

\end{abstract}


\section{Introduction}
Most massive stars are formed in clusters \citep{Lada2003} along with hundreds to thousands of intermediate and low-mass members \citep{Smith2009}. Typically, a massive star cluster may contain one to dozens of O-type stars as well as tens to hundreds of B-type stars.
The presence of massive stars in star clusters probably define their evolution, whether they maintain the group's gravitational ligature and contribute to a fast gas removal \citep{Vine2003,Goodwin2006,Pol2010}.\\
On the other hand, in compact (0.1-0.3 pc half-mass radii) clusters, most O-type stars can be ejected by dynamical interactions near the potential center \citep{Fujii2011,roman11,roman12,roman13,Oh2016}. 
In this case, we may speculate that B-type stars probably inherit the dominion of the gravitational potential and the dynamics of the cluster. 
This also should be the case when a star cluster does not form any O-type star and the occurrence of B stars is usually more common (e.g. 25Ori in the Orion OB1a sub-association, \citealt{Briceno2007}), with B-type stars taking the role of the O-stars, and becoming very good tracers for the cluster's initial mass function (IMF) \citep{Hohle2010}. This, in consequence, makes the study of B stars very important to understand the cluster's evolution.
Therefore, their correct identification and classification is crucial. \\
Spectral classification of intermediate to massive stars in the near-infrared (NIR) is advantageous, since it allows to study regions with high dust extinction, where most of them are still embedded.
Unfortunately, it has the disadvantage that the spectral features available in the NIR to classify hot (both O and B) stars are probably not enough to allow classification bellow one to two spectral sub-classes.
For this reason, the classification of this type of stars has been made preferentially using optical spectral features (e.g. \citealt{Hanson1996}), where we can find a large number of spectral lines. 
However, many past observations were done using low-resolution spectrographs, and considering the fact that massive stars usually show weak and only a few metal lines, the general spectral classification of B-type sources had not been precise so far. Generally, the classification systems in temperature scale, as well as in luminosity class, take into account lines of Helium (He {\sc i}, He {\sc ii}) and the characteristic lines of the Balmer series \citep{Gray2009, Sota2011}. The spectral features taken as base for the classification in the NIR are the lines of the Bracket series (Br11, Br13, Br15 and Br16 mainly).
Some works such as \citet{Hanson2005} and \citet{Blum1997}, present the NIR classification for OB stars, including only early B-type stars, by using some optical features for their study . Another example is the work of \citet{Liu2015}, where early and late types are included, with an automated method designed for cooler stars. \\
In this work we made use of NIR spectra from the Apache Point Observatory Galaxy Evolution Experiment (APOGEE2-N; see \citealt{Maje2017}) obtained during phase IV of the Sloan Digital Sky Survey (SDSS, \citealt{Blanton2017,Gunn2006}). We show that APOGEE2-N data can be used for spectral classification purposes of normal B-type stars. However, in a first stage it is necessary to complement the data of SDSS APOGEE2-N with optical spectra, which allow to make a detailed classification. This also permits to resolve the classification down to sub-classes from the classic features of the MK system.
In turn, this results would help to classify stars in other APOGEE2-N programs that lack optical counterparts in the future.
 For instance, we have a list of unclassified B-type stars in APOGEE2-N ancillary science program ``The W3/4/5 Super Star Forming Complexes: Unraveling Multiple Massive Star Populations'', on which we expect to apply the classification scheme presented in this paper. Moreover, having a precise classification scheme for intermediate mass stars opens the possibility to obtain their precise physical properties and to study them in the context of the early evolution of stellar clusters.\\
The primary goal of this work is to obtain a well defined relationship between optical and infrared spectra for B stars. We have a large sample (316 star spectra) to work with and the goal is to expand the classification sequence before exposed by \citet{Roman2018} to late B-type sources.
With this, we expect to be able to use the relation found for regions with B stars that do not have an optical counterpart for their classification, but have measurements in the NIR.\\
The paper is organized as follows: after this introduction, we describe our sample selection in Section \ref{s:sample} (optical and NIR). Later, we will present our classification method. For last, we expose our main results and the discussion.
 
\begin{figure}
    \centering
    \includegraphics[width=0.99\linewidth]{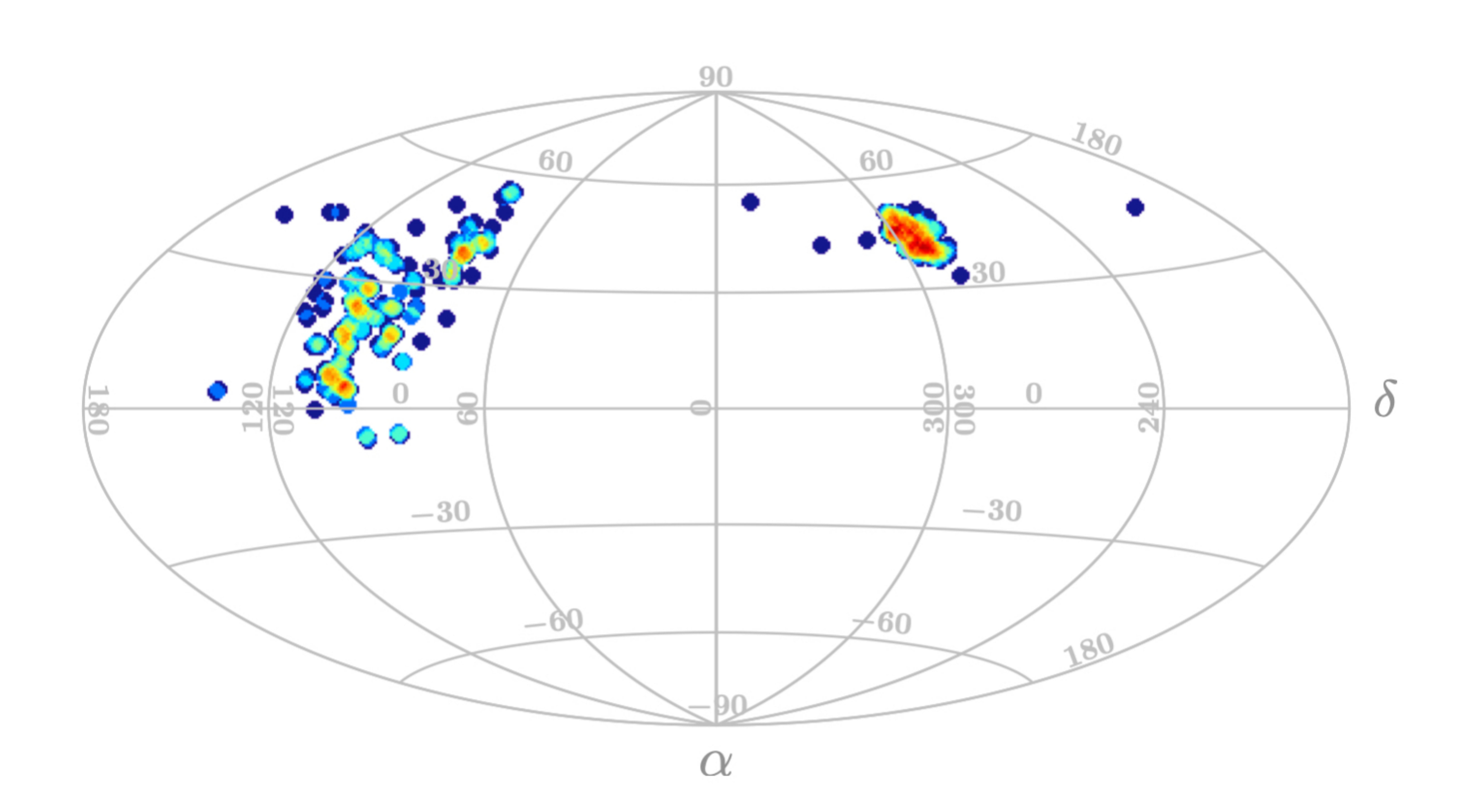}
    \caption{The panel shows the sky map positions, on equatorial coordinates, of the B-type stars in the selected sample.}
    \label{radec}
\end{figure}
\section{Sample \label{s:sample}}

\subsection{Near-infrared data}
We made use of NIR (H-band) spectra from APOGEE2-N program on its second phase \citep{Maje2017}.
The main scientific sample of APOGEE2-N is composed of red giant stars (for further information see \citealt{Holtzman2015,Nidever2015,Zamora2015,Garcia2016,Zasowski2017}) from all components of the Milky Way in order to reconstruct its formation history, chemical evolution and kinematics.
Using its two equivalent spectrographs \citep{Wilson2010} attached to 2.5m class telescopes, one on the Apache Point Observatory on New Mexico (APOGEE2-N) and the other at Las campanas observatory on Chile (APOGEE2-S), the survey is able to cover the entire Milky Way.
The spectrographs allow to obtain a large number of spectra from the Galaxy with a resolving power $R\sim$22500 covering the ranges 15145-15810 \AA\ (blue), 15860-16430 \AA\ (green), and 16480-16950 \AA\ (red).\\
The APOGEE-2 DR14 \citep{Abol2018} catalog contains a large number of non classified B-type stars that were used as telluric calibrators and, as shown by \citet{Roman2018}, it is possible to use APOGEE-2 spectra to classify massive stars. In the next sections we will show how this can be extended down to late B-type stars (intermediate mass sources).  

\begin{figure*}[htb]
	\centering
	\includegraphics[width=0.49\linewidth]{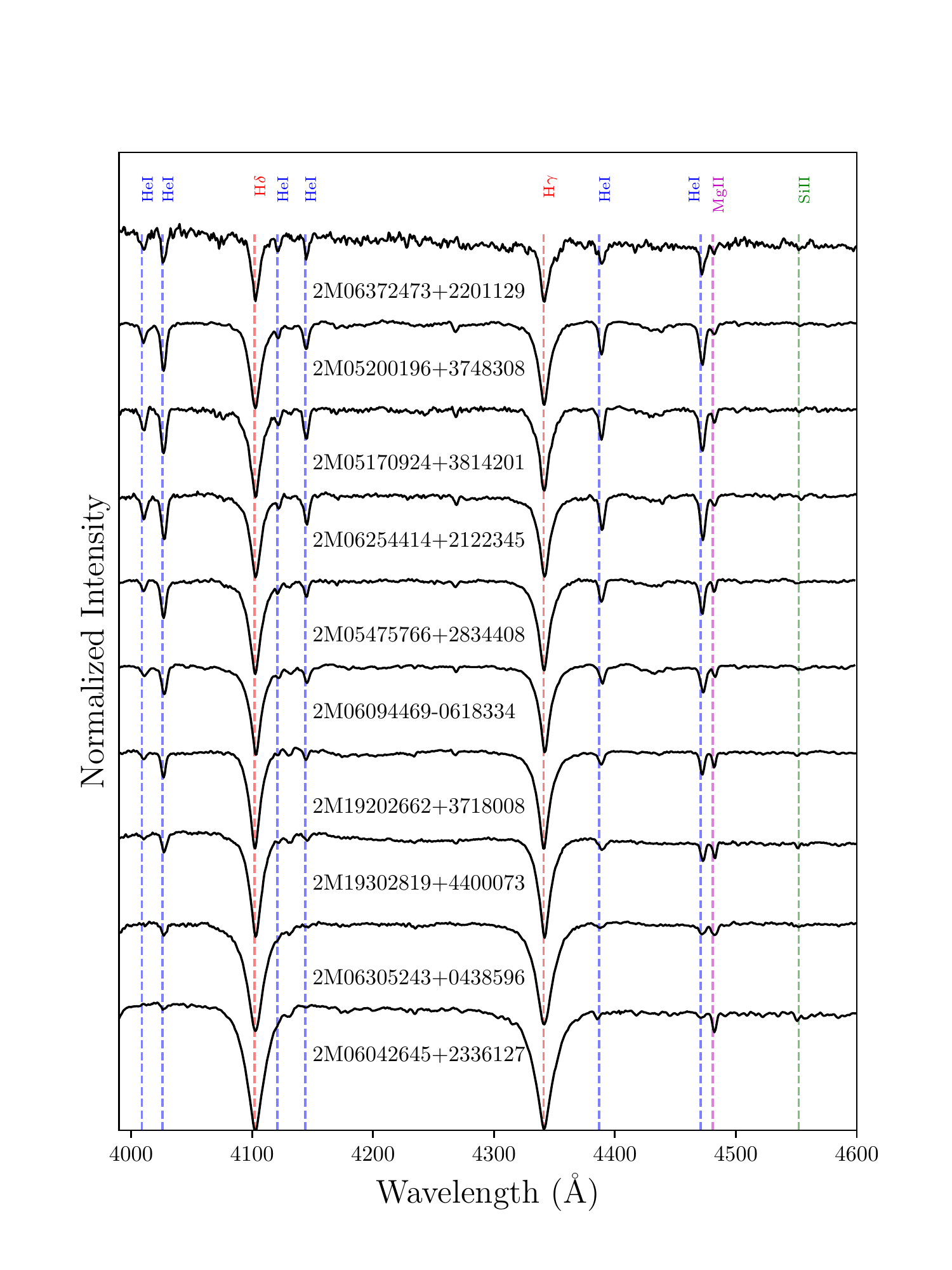}
	\includegraphics[width=0.49\linewidth]{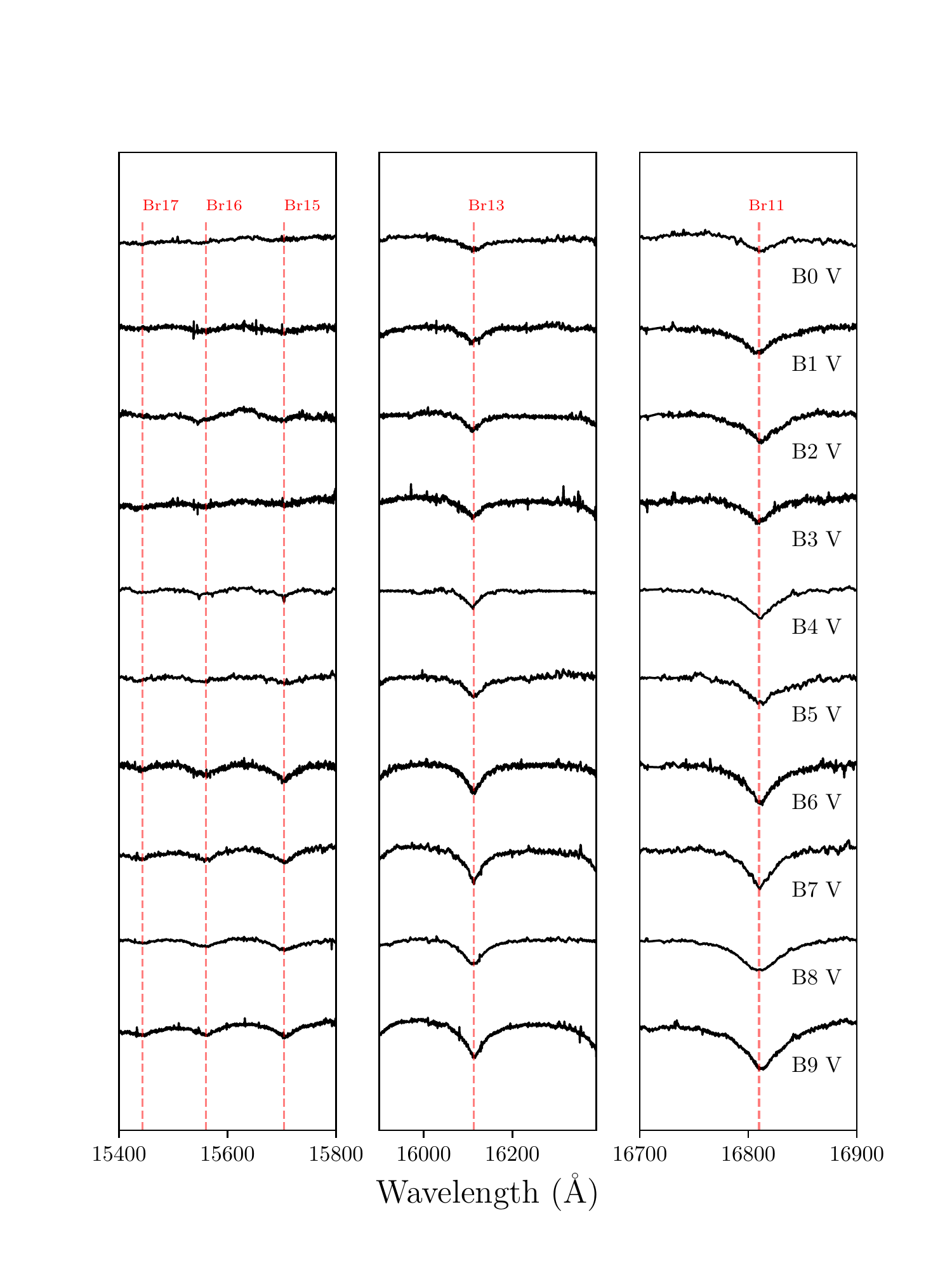}
	\caption{Examples of the LAMOST DR3 and SDSS-IV/APOGEE2 DR14 spectra used in this work. Left: the panel shows the principal features in the 
	4000\AA-4600\AA\ spectral range: He {\sc i} $\lambda$4009, He {\sc i} $\lambda$ 4026, H$\delta$ $\lambda$4102, He {\sc i} $\lambda$ 4121, He {\sc i} $
	\lambda$ 4144,  H$\gamma$ $\lambda$4341, He {\sc i} $\lambda$ 4387, He {\sc i} $\lambda$ 4471, Mg {\sc ii} $\lambda$ 4481 and Si {\sc iii} $\lambda$ 
	4552]. Right: the panel shows the APOGEE2-N spectra, in which we can see the Brackett series lines Br11 $\lambda$16811.111, Br13 $\lambda$16113.714, 
	Br15 $\lambda$15704.952, Br16 $\lambda$15560.699 and Br17 $\lambda$15443.139.}
	\label{spec-maps}
\end{figure*}

\subsection{Optical data}
As mentioned above, in order to calibrate and test our empirical spectral classification scheme, it is necessary to complement the NIR data of this study, with optical spectra.  
For this purpose, we made use of optical data from the Large Sky Area Multi-Object Fiber Spectroscopic Telescope (LAMOST, \citealt{Cui2012}).
LAMOST focuses on extragalactic sources, as well stars aiming to study the structure and evolution of the Galaxy. It uses a reflecting Schmidt telescope with 4000 fibers in a field of view of 20 deg$^2$. 
To date, 5.75 million spectra have been published divided on three data releases, which included 4.66 million high-quality spectra with SNR $\geq$ 10. 
The main LAMOST instrument provides spectra in the wavelength range of 3600-9100 \AA\  with a resolution of $\sim$1800 at $\sim$5500 \AA\.\\

The selection of the B-type stars sample used here was made by eye, based on a careful hand-picking inspection of the APOGEE2-N spectra. Then we cross-matched the resulting list with the LAMOST DR3 catalog. 
As we made use of APOGEE2-N spectra, the B-type stars sample lies on the Northern Hemisphere as shown in figure \ref{radec}. We obtained a total of 316 stars with optical counterparts and named them as ``control'' sample. It is important to mention that, in this sample, we did not include any obviously abnormal B-type spectrum, neither emission line stars.

\begin{figure*}
	\centering
	\includegraphics[width=0.9\linewidth]{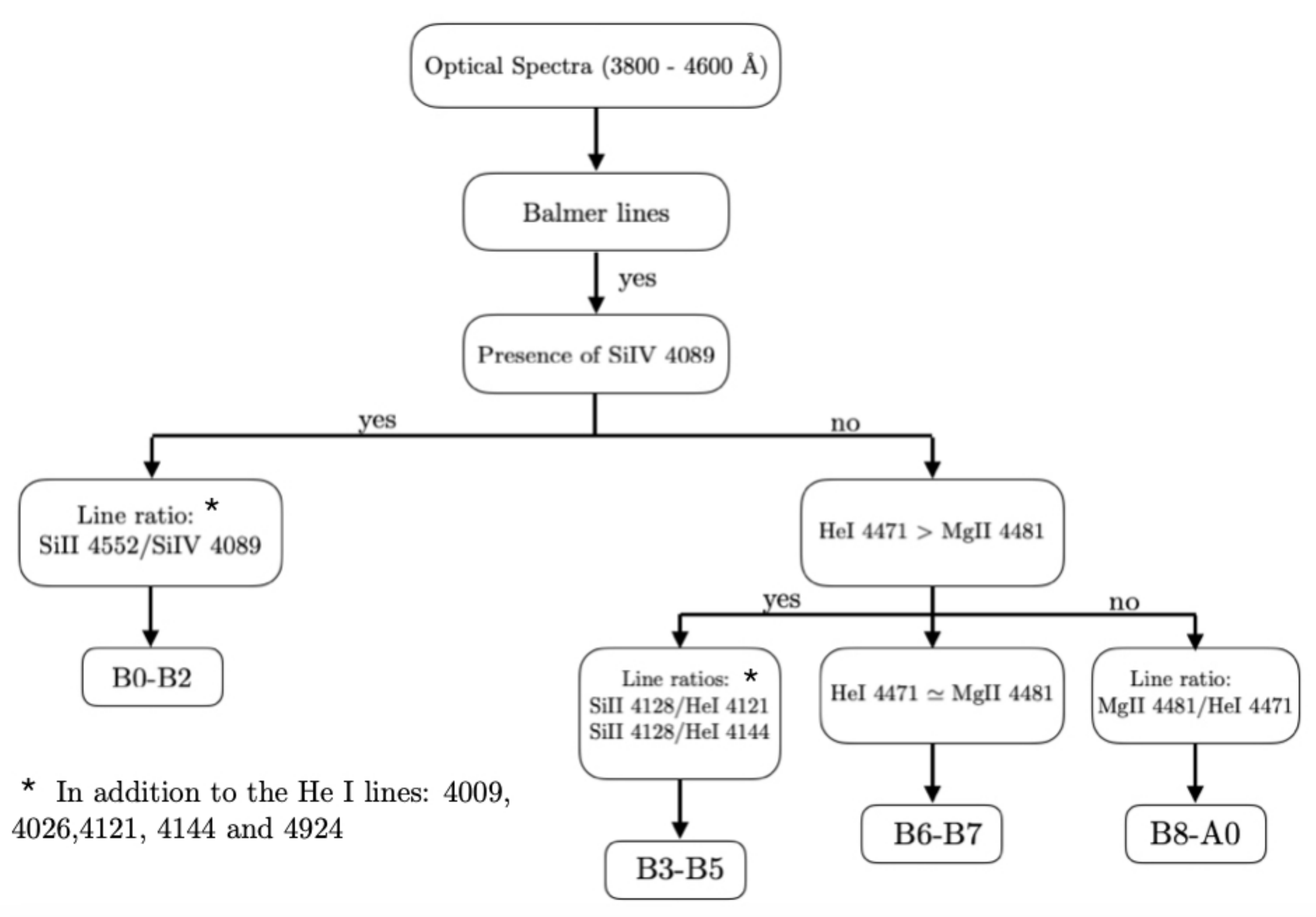}
	\caption{The scheme shows our B star classification process, starting from the spectral signatures present in the blue-optical spectra.}
	\label{class}
\end{figure*}

\section{Spectral classification\label{s:classification}}
\subsection{Optical Spectra\label{s:classification:ss:optical}}
The blue-optical spectra of normal B-type stars are dominated by the hydrogen absorption lines of the Balmer series and by He {\sc i} lines. The intensity of the Balmer lines increases towards later types, reaching its maximum for early A-type stars, while the intensity of the He {\sc i} lines increases with the temperature, reaching its maximum for stars of B2-B3 types \citep{Gray2009}. These spectral features were thus used to develop our methodology. \\
We derived the spectral types for the \textit{control} sample, using the spectral features present in the blue-optical (where we have a larger number of spectral lines and where the Morgan-Keenan spectral type system was originally defined). Then, we compared the results with the observed behavior in the APOGEE2-N spectra, using the spectral lines of the Bracket series, namely Br11 and B13. A set of selected examples of the blue-optical and NIR APOGEE-2 data for B-type stars (from B0 to B9) are shown in Figure \ref{spec-maps}.\\
During the classification procedure (see Figure \ref{class}), we choose to use the spectral lines in the 3800-4600 \AA\ range of the blue-optical window.
Using the task \textit{continuum} of the ONEDSPEC package of the Image Reduction and Analysis Facility (\texttt{IRAF\footnote{http://ast.noao.edu/data/software \label{refnote}}}) software, we rectified the continuum of the spectra, and we performed careful line parameter measurements with the \textit{fitprofs} task assuming Gaussian profiles for all spectral lines.\\ 
After the line measurements, we followed the prescriptions by \citet{Morgan1943}, \citet{Gray2009} and \citet{Sota2011} to do the classification from the optical spectra:
Fist, we considered that the presence (or absence) of the Si {\sc iv} $\lambda$4089 line indicate earlier spectral types, with observed intensities correlating with the increase or decrease of the effective temperature of the star.  
For those B-type spectra presenting this feature, we used the line ratio Si {\sc ii} $\lambda$4552 / Si {\sc iv} $\lambda$4089  for the earliest \textit {B0-B2} types.
In the absence of the mentioned line, we used instead the He {\sc i} $\lambda$4471 and Mg {\sc ii} $\lambda$4481 lines. 
For those cases where the He {\sc i} lines are stronger than the Mg {\sc ii} lines, we used the Si {\sc ii} $\lambda$4128 / He {\sc i} $\lambda$4121 and Si {\sc ii} $\lambda$4128 / He {\sc i} $\lambda$4144 line ratios, which are useful to classify B-stars in the intermediate spectral types \textit{B3-B5}. Also, we took into account the intensity of the He {\sc i} lines  $\lambda$4009, $\lambda$4024, $\lambda$4121, $\lambda$4144 and $\lambda$4924 for the correct identification of the earlier types B0-B5.\\
In those cases for which the He {\sc i} $\lambda$4471 has a similar intensity to the Mg {\sc ii}, we assigned spectral types \textit {B6-B7}. In the opposite case, when the Mg {\sc ii} line is stronger than that He {\sc i} $\lambda$4471, we used the Mg {\sc ii} $\lambda$4481 / He {\sc i} $\lambda$4471 line ratio, which grows at later types and denotes spectral types \textit{B8-A0}. \\
Finally, it is useful to notice that all mentioned lines are sensitive to the luminosity class of the star \citep{Gray2009}, and as we move to later spectral types, there are cases in which an earlier B-type star with weak Helium lines might be confused with a later type source. To prevent this, we also considered the luminosity classes in the classification process.
For this purpose, we used the intensities of the hydrogen Balmer lines H $\gamma$ and H $\delta$ as luminosity indicators. 
In addition, the presence of prominent N {\sc ii} $\lambda$3995, Si {\sc ii} $\lambda$4552 and the He {\sc i} $\lambda$4929 line features was considered to distinguish supergiants and giants from main sequence stars.\\
Once we classified the \textit{control} sample stars in the blue-optical window, the next step was to study the associate behaviors in the NIR spectral range counterpart. \\

\begin{figure*}[htb]
	\centering
	\includegraphics[width=0.45\linewidth]{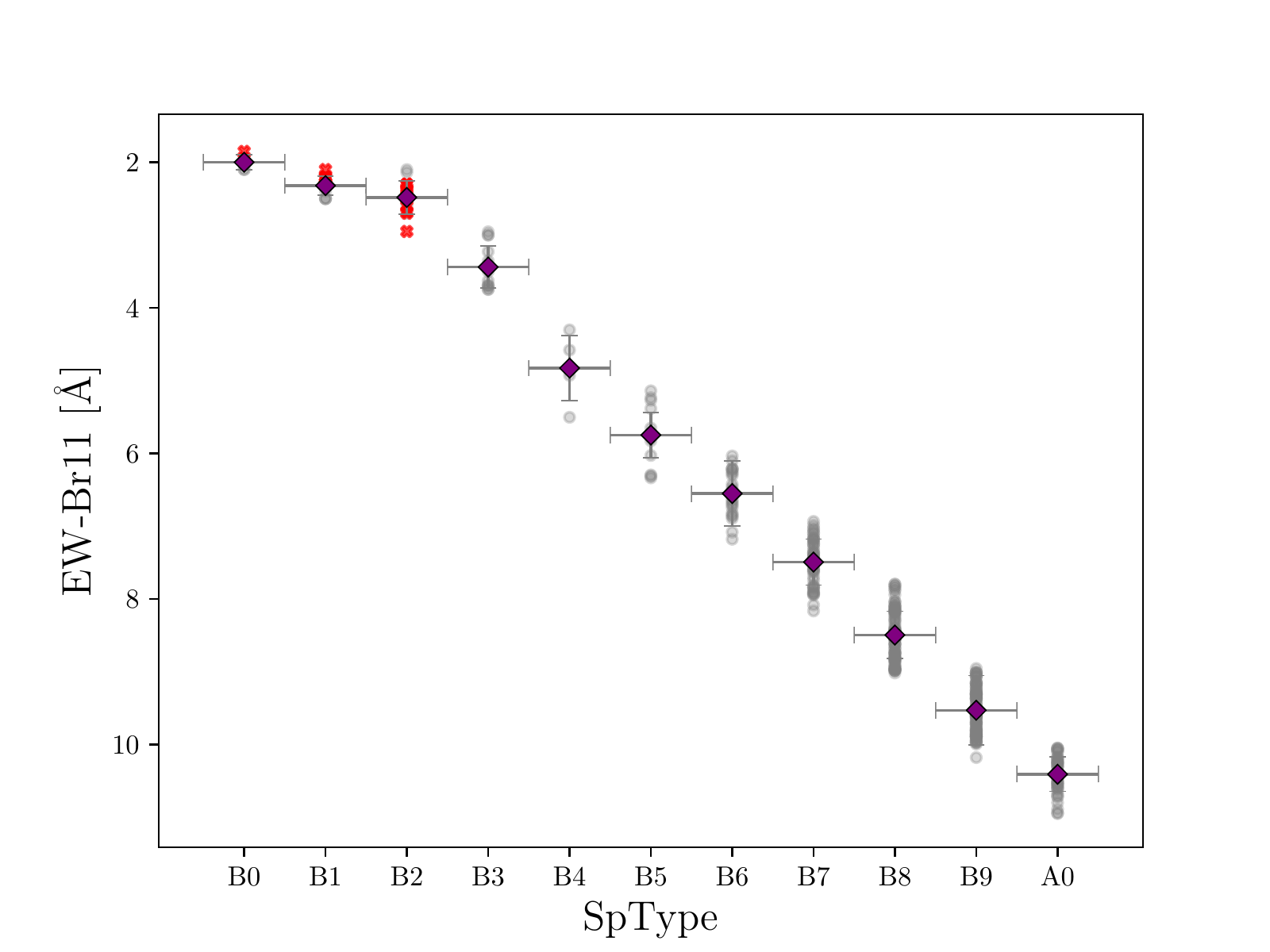}
	\includegraphics[width=0.45\linewidth]{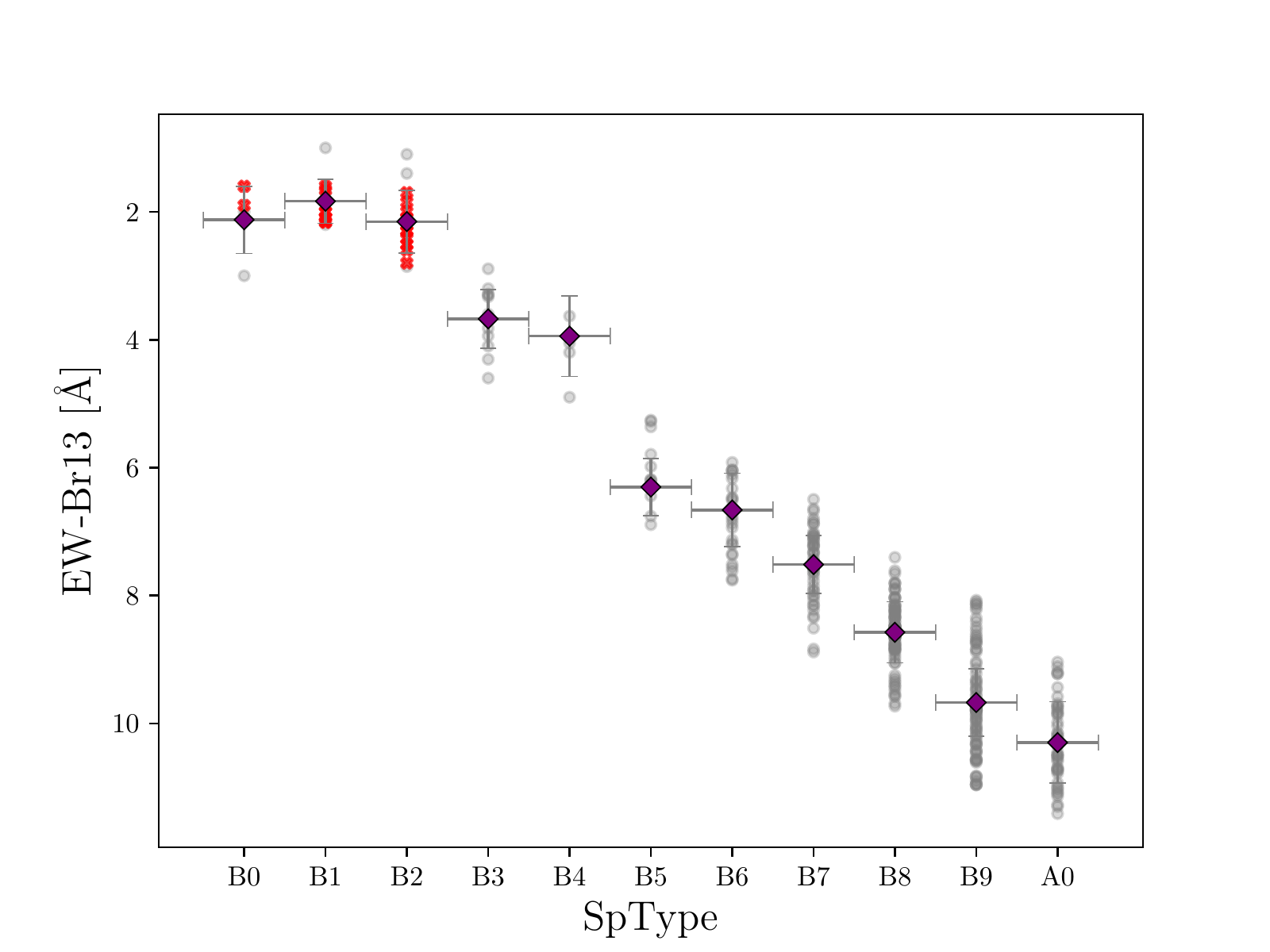}
	\caption{In the left and right panel we show, respectively the Br11 and Br13 EW values versus the spectral type obtained from the classification scheme. The purple dots represent the median of each type with its respective error bar (corresponding to one spectral subtype on the x-axis and the respective standard deviation on the y-axis).  The red x symbol represent the early sources (B0-B2) we included from the work of \citet{Roman2018}. The full sample is represented in the gray circles.}
	\label{relation}
\end{figure*}

\subsection{Near-infrared Spectra\label{s:classification:ss:optical}}
For the NIR APOGEE-2 spectra of our \textit{control} sample, we used the same \texttt{IRAF} tasks/procedures described above, to rectify the continuum and to perform line measurements.
For this part of our work, we selected the spectral window between 15400\AA\ and 16900\AA. With the task \textit{fitprofs}, we measured the equivalent widths (EW) and full width half maximum (FWHM) of the Brackett 11 $\lambda$16811.111 (Br11) and the Brackett 13 $\lambda$16113.714 (Br13) spectral lines. Then we determined the dependence of the NIR EW and the spectral types obtained from the blue-optical window. In Table 1 we present the line parameters measurements for all stars in our APOGEE2-N \textit{control} sample.\\
The associated EW errors come from the Gaussian profile fitting made for the Brackett lines measurement, and were computed with the \textit{fitprofs} error tool. This routine uses an error estimator based on Poisson statistics  determined from the fit of a Gaussian function to the spectral line. The main parameters for the estimator are the standard deviation associated to the best-fit Gaussian, and the level of external noise (\textit{sigma0} and \textit{invgain}, respectively). At first approximation, a value of \textit{invgain} = 0 is enough if the signal to noise value is high. The \textit{sigma0} parameter is estimated from the root mean square value measured in the continuum. This way, the average error associated with each measurement is approximately 0.5 \AA, with small variations that depend on the signal to noise level of each spectrum. We considered one spectral subtype error on our B star classification scheme. This error comes from the optical classification presented on Figure \ref{class}.

\startlongtable
\begin{center}
\begin{deluxetable*}{lcccccccc}
\tabletypesize{\scriptsize}
\tablecolumns{8}
\tablewidth{0pt}
\tablecaption{NIR parameters for the control sample in APOGEE-2.}
\tablehead{
\colhead{ID} &
\colhead{R.A.} &
\colhead{Decl.} &
\colhead{Br11} &
\colhead{Br11} &
\colhead{Br13} &
\colhead{Br13} &
\colhead{Sptype} &
\\
\colhead{} &
\colhead{} &
\colhead{} &
\colhead{EW (\AA)\tablenotemark{a}} &
\colhead{FWHM (\AA)\tablenotemark{a}} &
\colhead{EW (\AA)\tablenotemark{a}} &
\colhead{FWHM (\AA)\tablenotemark{a}} &
}
\startdata
 2M06094469-0618334 & 06:09:44.69 & -06:18:33.47 & 5.356(0.1457) & 28.16(1.496) & 5.94(0.08774) & 16.52(1.929)	& B5.V \\ 
 2M19202662+3718008 & 19:20:26.62 & +37:18:00.84 & 6.378(0.5435) & 37.07(2.448) & 6.927(0.6325) & 53.42(3.777) & B6.V \\	 
 2M19302819+4400073 & 19:30:28.19 & +44:00:07.37 & 7.217(0.2332) & 23.41(2.966) & 7.326(0.4639) & 41.67(2.673) & B7.V \\
 2M06305243+0438596 & 06:30:52.43 & +04:38:59.64 & 8.347(0.8675) & 38.72(7.946) & 8.736(0.3764) & 65.61(0.7901) & B8.V \\
 2M06042645+2336127 & 06:04:26.45 & +23:36:12.74 & 9.384(0.2653) & 15.65(3.371) & 8.709(0.5349) & 44.83(3.193)	 & B9.V \\
 \enddata
 \tablenotetext{a}{
The values between parenthesis represents the errors on each parameter respectively. The full version of this table will be available in the electronic version of this paper.}
 \label{tabla1}
\end{deluxetable*}
\end{center}

 \section{Results \label{s:results}}
Using the methodology described in the previous section, we classified 316 spectra of the \textit{control} sample, from which about the 90\% corresponds to B-type stars.


In the left panel of Figure \ref{relation} we show the observed relation between the EWs of the Br11 line, and the spectral types of the B-type stars on our sample, as determined from the blue-optical spectra. The plot shows that the Br11 EW changes linearly with the spectral types, with the hydrogen lines becoming deeper and less intense as the temperature of the star rises. In the right panel of the same figure, a similar relationship is shown for the Br13 transition. We can notice a similar linear relation, although with larger dispersion, mainly for the later and cooler B-types. This is probably because the profile of the Br13 line sometimes appears slightly broadened by what we suspect to be an unknown, unresolved absorption line, possibly resulting from the helium transitions associated to a multiplet at $\sim \lambda$16105\AA. 

\begin{figure}
	\centering
	\includegraphics[width=0.95\linewidth]{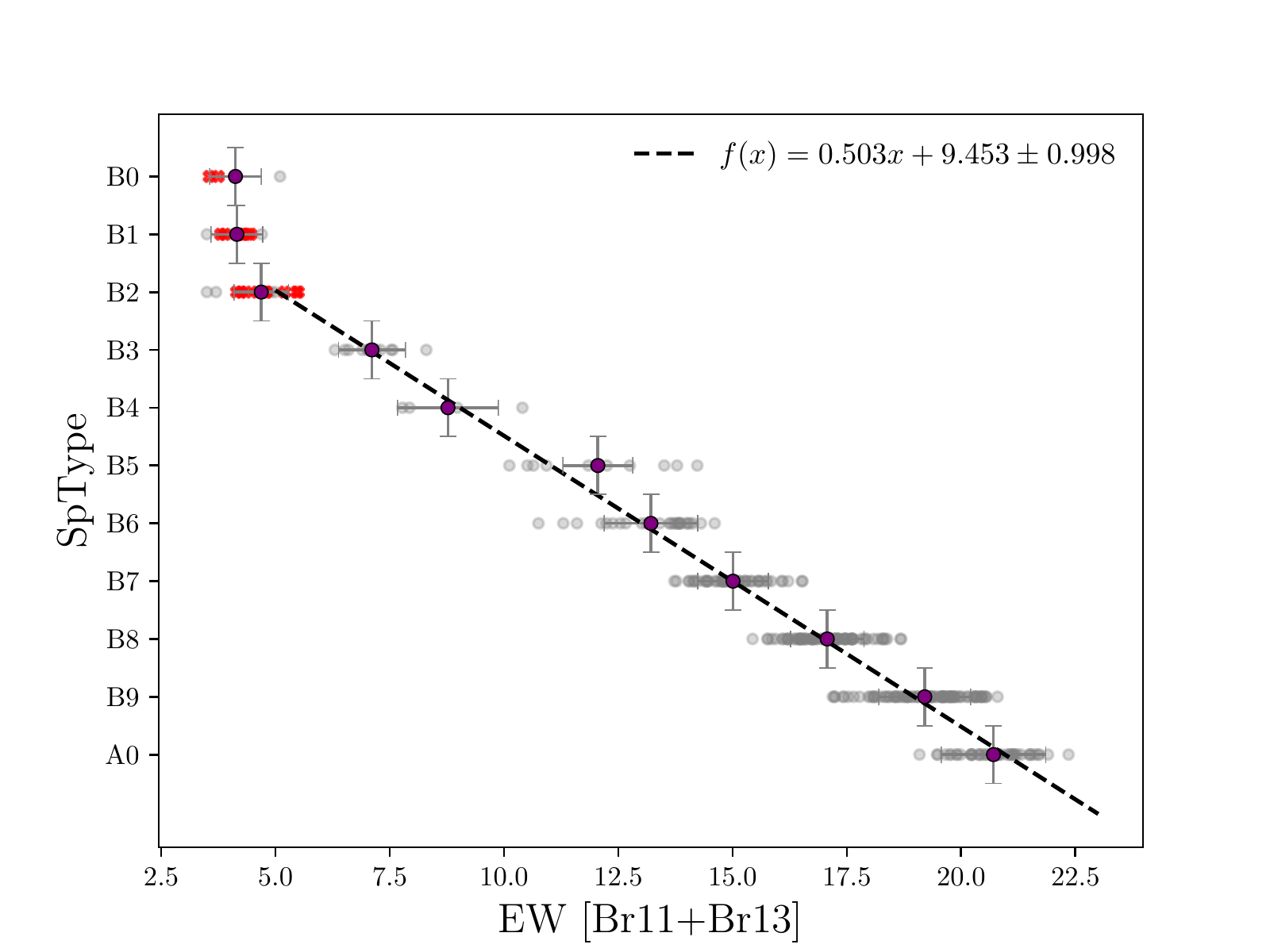}
	\caption{The figure shows the sum of the Br11 and Br13 EW values versus the spectral type obtained from the classification. The dotted line represent the NIR relation found for the B-type stars. The purple points represent the median per bin of spectral type, and the gray color indicate the full sample in comparison. The red x symbol represent the early sources (B0-B2) we included from the work of \citet{Roman2018}.}
	\label{sum}
\end{figure}

In Figure \ref{sum}, we show the observed relation between the sum EW[Br11+Br13] and the optical spectral types. As expected, the linear relation found for stars in the temperature range of the B3 to A0 types is maintained, with the observed behavior deviating from the linear trend in case of B-star hotter that B1-B2, an effect also noticed by \citet{Roman2018}. 

We use a jackknife resampling procedure to obtain the most probables values for the median and the correlation for the linear fitting in Figure \ref{sum}. This fitting was weighted by their respective errors and only took account those stars with espectral types B3-A0. This way, the respective relation between the spectral type and the EW[Br11+Br13] is:

\begin{equation}
SpType = 0.503EW[Br11+Br13]+9.453 \ (\pm0.998)
\end{equation}

This equation associates a numerical value for the spectral type with the Br11 and Br13 EWs and works satisfactorily for B3 to A0 spectral types, and is consistent with the previous work of \citet{Roman2018}. With this equation we can provide spectral type estimates for normal B-type stars with no optical counterpart data. \\

\begin{figure}
    \centering
    \includegraphics[width=0.95\linewidth]{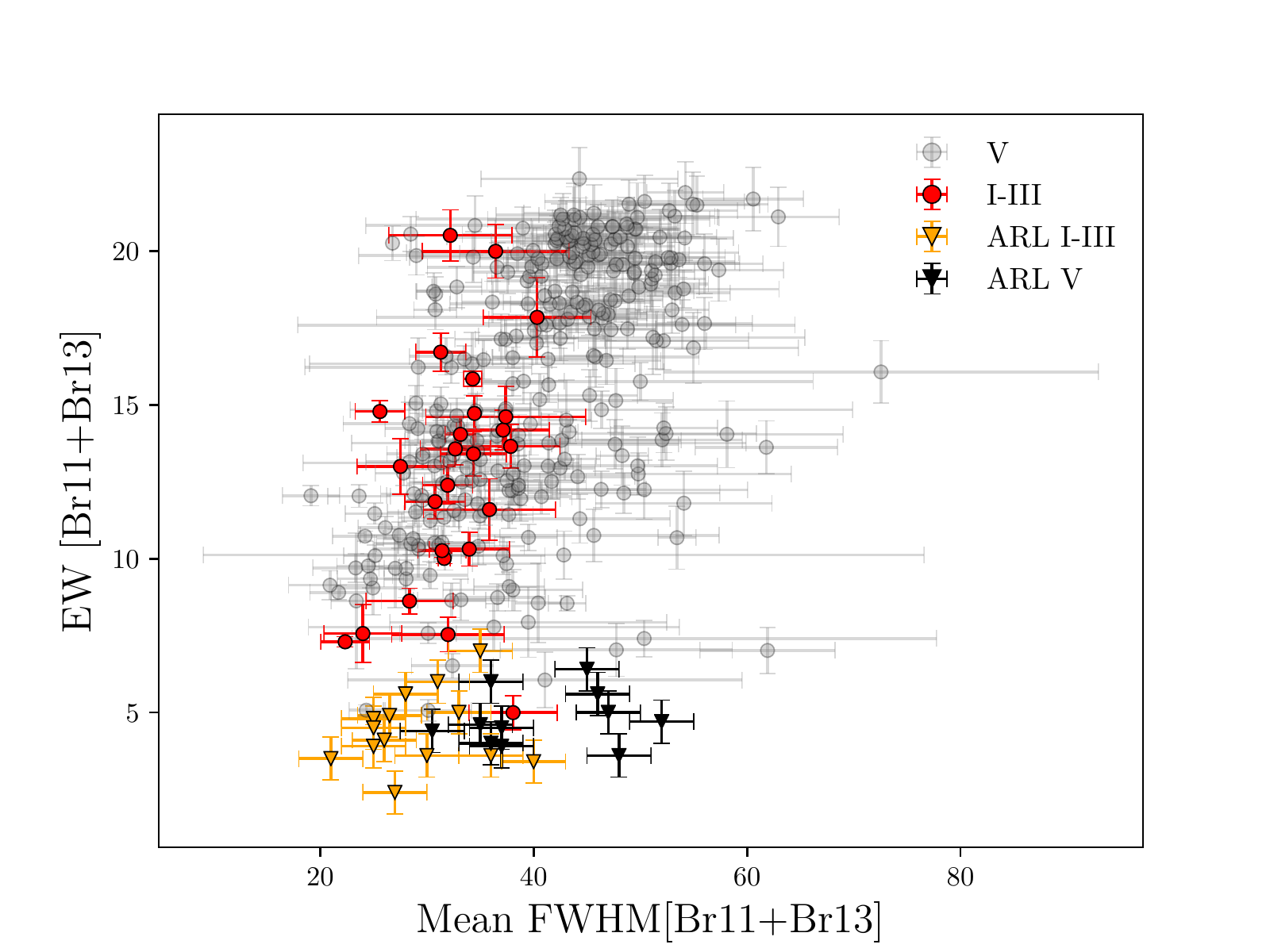}
    \caption{The panel shows the full width at half maximum of the control sample. The red circles correspond to I-III classes and the gray circles correspond to class V stars. The orange (classes I-III) and gray (class V) triangles belongs to the work of \citet{Roman2018} (ARL).}
    \label{fwhm}
\end{figure}

Figure \ref{fwhm} shows the relation between the sum of the EWs of the Br11 and Br13 lines and their associated FWHM. For the \textit{control} B-type sample, most stars are of  luminosity class {\sc v}, with 25 exemplars belonging to the supergiant and giants classes. 
There is a clear separation between the classes, for the latest types. For the earliest types, we can see how classes tend to overlap, as the sum of their EW decrease, because most of the hydrogen atoms are ionized at higher temperatures.\\

\begin{figure}
    \centering
    \centerline{\includegraphics[width=1.2\linewidth]{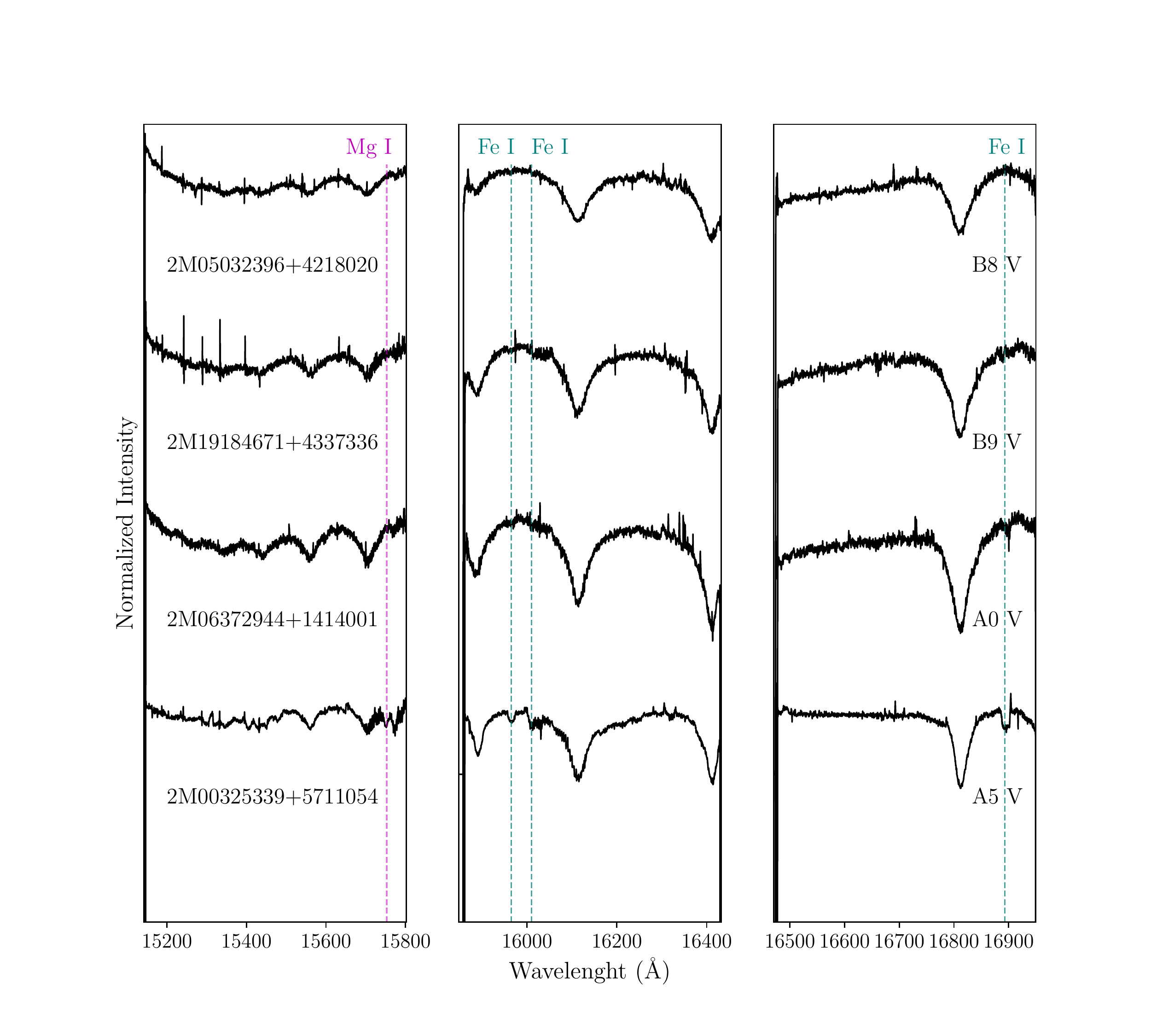}}
    \caption{The panel shows the transition between B8-A5 spectral types. We can see how the spectral lines of metallic elements (Mg {\sc I} $\lambda$15753, Fe {\sc I} $\lambda$15965, Fe {\sc I} $\lambda$16010 and Fe {\sc I} $\lambda$16893) begin to appear as we move to later spectral types \citep{Roman20}.}
    \label{tran}
\end{figure}

When classifying B stars, we usually look for specific spectral features suggested in the literature, such as the He {\sc I} lines and metallic lines we mentioned in Section \ref{s:classification}. As mentioned before, we can clearly distinguish B from O stars due to the absence of He {\sc II} in their spectra. Correspondingly we can distinguish B from A stars by the absence of He {\sc I} lines (they are almost imperceptible, or totally absent). In addition, the growth of the hydrogen lines from B9 to later types seems to be smooth and rapid \citep{Morgan1943} reaching its maximum at A2 \citep{Liu2019}. This may represent a problem when distinguishing B spectral type from A stars. The indicators we have for those cases are the absence of He {\sc I} lines and the evident growth of metallic lines (K lines are strong indicatives, \citealt{Gray2009}). \\ According to \citet{Morgan1943} it is possible to discriminate these spectral types through their absolute magnitudes or their color indices. \\
All of these features, however, work for those cases for which we have optical spectra counterparts.
For our semi-empirical NIR sequence, we can see from the work of \citet{Roman2018}, that for later spectral types, some metallic lines begin to appear as an additional indicator, while the Brackett lines continue growing as a function of spectral type reaching a maximum at A2. \\
In Figure \ref{tran} we compare APOGEE spectra of late B and early A main sequence stars. In this figure, we notice the metallic spectral features Mg {\sc I} $\lambda$15753, Fe {\sc I} $\lambda$15965, Fe {\sc I} $\lambda$16010 and Fe {\sc I} $\lambda$16893 standing out from early to late types \citep{Roman20}, while the Brackett lines Br11 and B13 increase in depth. These features, together, allow us to distinguish between both spectral types.  


\section{Discussion and Summary \label{s:discussion}}
In the previous sections we have described a classification method for B-type stars based on the analysis of a robust sample of APOGEE NIR and LAMOST optical spectra. We have shown that our semi-empirical method is reliable for spectral classification of B-type stars down to one spectral subtype, in a similar way to the work of \citet{Roman2018}.\\
We showed that there is a linear relationship between the EW of the Brackett lines (Br11 and Br13) present in the H-band window of the APOGEE-2 spectra, and the spectral type for most B stars, as previously suggested by \citet{Hanson1996}. 
On that study, \citeauthor{Hanson1996} used NIR K-band (2 $\mu m$) spectra of hot stars and found that the EWs of the Brackett lines series increase in case of later B types. However due to their smaller sample and the relatively low signal-to-noise of the spectra used on their work, at that time it was not possible to properly distinguish between early dwarfs and late supergiants stars. 
The EW x SpType NIR relation we found is well suitable to classify normal stars of spectral types in the range B3 to A0. From the work of \citet{Roman2018} and our results, we can confirm that there is a change on the slope of the NIR relation, for the earliest B star types (B0-B2). The tendency in the sequence of \citet{Roman2018} starts to change from O7-O9 types, and then changes again in the B0-B3 range, when the EWs begins to grow. 
The NIR relation found on our work, probably extends down to mid-to late A types, but a complete study down to later A types is beyond the scope of this paper. 
\\
As for the FWHM, from Figure \ref{fwhm}, we can separate our \textit{control} sample of B-type stars by luminosity classes for the latest types. This tendency seems to disappear as we move to the earlier types as they overlap. Our data seems to show a separation between supergiants and giants from main sequence stars, but as our sample is mostly composed by late spectral type stars with only a few high luminosity sources, we can not be entirely conclusive on that regard. Also, we include the earliest sources before presented by \citet{Roman2018} to make a connection between the two studies, were we can see that there is a correspondency between their results and ours.\\
As future work, we plan to use the NIR relation found in this study in star formation complexes containing normal B-type stars, for which there are no optical spectra available to apply the ``classical'' classification scheme. Therefore, the main advantage of our classification scheme is that it can be very useful in the spectral classification of young B-type stars still heavily embedded in their parental molecular clouds.

\acknowledgments
\begin{center}
ACKNOWLEDGMENTS.
\end{center}
VRP acknowledge support from a graduate studies fellowship from CONACYT/UNAM Mexico. VRP, CRZ and JH acknowledges support from UNAM-DGAPA-PAPIIT grants IN108117 and IA102319, Mexico. ARL acknowledges financial support provided in Chile by Comisi\'on Nacional de Investigaci\'on Cient\'ifica y Tecnol\'ogica (CONICYT) through the FONDECYT project 1170476 and by the QUIMAL project 130001, and support in Mexico from the PREI DGAPA UNAM program for academic exchange scholarship.
DAGH acknowledges support from the State Research Agency (AEI) of the Spanish Ministry of Science, Innovation and Universities (MCIU) and the European Regional Development Fund (FEDER) under grant AYA2017-88254-P.\\

Funding for the Sloan Digital Sky Survey IV has been provided by the Alfred P. Sloan Foundation, the U.S. Department of Energy Office of Science, and the Participating Institutions. SDSS-IV acknowledges
support and resources from the Center for High-Performance Computing at
the University of Utah. The SDSS web site is www.sdss.org.

SDSS-IV is managed by the Astrophysical Research Consortium for the 
Participating Institutions of the SDSS Collaboration including the 
Brazilian Participation Group, the Carnegie Institution for Science, 
Carnegie Mellon University, the Chilean Participation Group, the French Participation Group, Harvard-Smithsonian Center for Astrophysics, 
Instituto de Astrof\'isica de Canarias, The Johns Hopkins University, Kavli Institute for the Physics and Mathematics of the Universe (IPMU) / 
University of Tokyo, the Korean Participation Group, Lawrence Berkeley National Laboratory, 
Leibniz Institut f\"ur Astrophysik Potsdam (AIP),  
Max-Planck-Institut f\"ur Astronomie (MPIA Heidelberg), 
Max-Planck-Institut f\"ur Astrophysik (MPA Garching), 
Max-Planck-Institut f\"ur Extraterrestrische Physik (MPE), 
National Astronomical Observatories of China, New Mexico State University, 
New York University, University of Notre Dame, 
Observat\'ario Nacional / MCTI, The Ohio State University, 
Pennsylvania State University, Shanghai Astronomical Observatory, 
United Kingdom Participation Group,
Universidad Nacional Aut\'onoma de M\'exico, University of Arizona, 
University of Colorado Boulder, University of Oxford, University of Portsmouth, 
University of Utah, University of Virginia, University of Washington, University of Wisconsin, 
Vanderbilt University, and Yale University.

Guoshoujing Telescope (the Large Sky Area Multi-Object Fiber Spectroscopic Telescope LAMOST) is a National Major Scientific Project built by the Chinese Academy of Sciences. Funding for the project has been provided by the National Development and Reform Commission. LAMOST is operated and managed by the National Astronomical Observatories, Chinese Academy of Sciences.


\software{TopCat \citep{topcat}, \texttt{IRAF} \citep{Tody1986,Tody1993}}

\bibliographystyle{yahapj}

\end{document}